\documentclass[aps,prb,reprint,amssymb,amsmath,showpacs,superscriptaddress,floatfix]{revtex4-1}

\usepackage{graphicx}
\usepackage{epstopdf}
\usepackage[english]{babel}
\bibpunct{[}{]}{,}{n}{}{}

\begin{document}

\title{Spin-orbit splitting of the conduction band in HgTe quantum wells: role of different mechanisms}

\author{G.\,M.~Minkov}
\affiliation{School of Natural Sciences and Mathemxatics, Ural Federal University,
620002 Ekaterinburg, Russia}

\affiliation{M.N. Miheev Institute of Metal Physics of Ural Branch of Russian Academy of Sciences, 620137 Ekaterinburg, Russia}

\author{V.\,Ya.~Aleshkin}
\affiliation{Institute for Physics of Microstructures  RAS, 603950 Nizhny Novgorod, Russia}

\author{O.\,E.~Rut}
\affiliation{School of Natural Sciences and Mathematics, Ural Federal University,
620002 Ekaterinburg, Russia}

\author{A.\,A.~Sherstobitov}

\affiliation{School of Natural Sciences and Mathematics, Ural Federal University,
620002 Ekaterinburg, Russia}

\affiliation{M.N. Miheev Institute of Metal Physics of Ural Branch of Russian Academy of Sciences, 620137 Ekaterinburg, Russia}

\author{A.\,V.~Germanenko}

\affiliation{School of Natural Sciences and Mathematics, Ural Federal University,
620002 Ekaterinburg, Russia}

\author{S.\,A.~Dvoretski}

\affiliation{Institute of Semiconductor Physics RAS, 630090
Novosibirsk, Russia}

\author{N.\,N.~Mikhailov}

\affiliation{Institute of Semiconductor Physics RAS, 630090
Novosibirsk, Russia}
\affiliation{Novosibirsk State University, Novosibirsk 630090, Russia}

\date{\today}

\begin{abstract}
Spin-orbit splitting of  conduction band in HgTe quantum wells was studied experimentally. In order to recognize the role of different mechanisms, we carried out detailed measurements of the Shubnikov-de Haas oscillations in gated structures with a quantum well widths from $8$ to $18$~nm over a wide range of electron density. With increasing electron density controlled by the gate voltage, splitting of the maximum of the Fourier spectrum $ f_0 $ into two components $ f_1 $ and $ f_2 $ and the appearance of the low-frequency component $ f_3 $ was observed. Analysis of these results shows that the components $f_1$ and $f_2$ give the electron densities $n_1$ and $n_2$ in spin-orbit split subbands while the $f_3$ component   results from magneto-intersubband oscillations so that $f_3=f_1 - f_2$.  Comparison of these data with results of self-consistent calculations carried out within the framework of four-band \emph{kP}-model shows that a main contribution to spin-orbit splitting  comes from the Bychkov-Rashba effect. Contribution of the interface inversion asymmetry to the splitting of the conduction band turns out to be four-to-five times less than that for the valence band in the same structures.
\end{abstract}


\maketitle

\section{Introduction}
\label{sec:intr}
The energy spectrum of semiconductors, which contain heavy elements, is determined to a significant degree by the spin-orbit (SO) interaction. In bulk semiconductors, the SO interaction in the crystal field, as a rule, is directly included in the Hamiltonian. Just this interaction determines the splitting of the $\Gamma_{15}$ valence band in semiconductors A$_3$B$_5$, A$_2$B$_6$, etc., and, thus, the energy distance between the bands $\Gamma_8$ and $\Gamma_7$ for zero quasimomentum value, $k = 0$. The absence of an inversion center in these materials leads to an additional splitting for $k\neq 0$.

In two-dimensional (2D) systems, other mechanisms of SO interaction become important. The first one is caused by a structure inverse asymmetry  resulting from different barrier heights, the asymmetry of doping, the gradient of the composition in the quantum well (QW) in the case when it consists of a multicomponent semiconductor. The second mechanism is the interface inverse asymmetry (IIA), and finally the potential gradient in a direction perpendicular to the 2D plane can play a crucial role. The latter mechanism, known as Bychkov-Rashba (BR) effect,  is of particular interest, since it can be controlled in the gated  structures.

Experimental studies of the valence band spectrum in HgTe quantum wells \cite{Minkov16,Minkov17} show that the valence band is strongly split  in the structures with different QW width ($d$); both for $d$ close to the critical width $d_c\simeq 6.3$~nm at which the massless Dirac-like spectrum is realized and for $d>8$~nm, where additional maxima at $ k\neq 0 $ turn out to lie higher in the energy than the hole states at $k=0$. The calculations  of spectrum with taking into account IIA \cite{Minkov16} show  reasonable agreement with the experimental results on the splitting of the top of the valence band. Such calculations also predict a large splitting of the conduction band at a low electron density. However, in structures with $d$ close to $d_c$ no noticeable splitting was experimentally  observed for $n<3.5 \times 10^{11}$~cm$^{-2}$. Spin-orbit splitting of the conduction band was observed and studied mainly in wide quantum wells, $d=(10-20)$~nm, for relatively high electron density,$n \approx 10^{12}$~cm$^{-2}$,  and was interpreted as a result of the BR effect \cite{Schultz96,Zhang01,Zhang02,Gui04}. Thus, the role of various mechanisms for the SO splitting of the  conduction band and their dependence on the QW width and the electron density remains unclear yet.

In this paper, in order to determine the role of different mechanisms of SO interaction, we carried out detailed measurements of the Shubnikov-de Haas (SdH) oscillations in structures with the quantum well width from $8$ to $18$~nm over a wide range of electron density. We observed appearance of three components in the Fourier spectra with the frequencies $f_1$, $f_2$, and $f_3$  when the electron density increases  with the increasing gate voltage ($V_g$). An analysis  of these results shows that the $f_1$ and $f_2$ components  give the electron densities $n_1$ and $n_2$ in the SO split subbands while the $f_3$ component  results from magneto-intersubband oscillations and it is equal to $f_3=f_1 - f_2$. The  SO splitting is quantified here as $\Delta n/(n_1+n_2)$, where $\Delta n=n_1-n_2$, can be experimentally found both as $(f_1-f_2)/(f_1+f_2)$ and as $f_3/(f_1+f_2)$. Comparison of these results with the self-consistent calculations carried out within the four-band \emph{kP} model shows that the  main contribution to the SO splitting comes from the BR effect while the IIA contribution is less than $(10-15)$ percent.

\section{Experiment}
\label{sec:expdet}

The HgTe quantum wells were realized on the basis of HgTe/Cd$_{x}$Hg$_{1-x}$Te ($x=0.39-0.6$)  heterostructure grown by molecular beam epitaxy on GaAs substrate with the ($013$), ($211$), and ($100$) surface orientation \cite{Mikhailov06}. The nominal QW widths were $d = (8.3-18)$~nm. The samples were mesa etched into standard Hall bars of $0.5$~mm width with the distance between the potential probes of $0.5$~mm. To change and control the electron densities  in the quantum well, the field-effect transistors were fabricated with parylene as an insulator and aluminium as a gate electrode. The measurements were performed at the temperature  $T=(4 - 12)$~K in the  magnetic field up to $6.0$~T. For each heterostructure, several samples were fabricated and studied. The main parameters are listed in  the Table~\ref{tab1}.

\begin{table}
\caption{The parameters of  heterostructures under study}
\label{tab1}
\begin{ruledtabular}
\begin{tabular}{ccccc}
number & structure & $d$ (nm) & $x$ & substrate orientation \\
\colrule
  1& 100623 & 18    & $0.39$   & (100)  \\
  2& 091228 &  14      & $0.60$    & (211)   \\
  3& 150224 & 10     & $0.52$   & (013)  \\
  4& 150629 & 8.3     & $0.58$   & (013) \\
\end{tabular}
\end{ruledtabular}
\end{table}

\section{Results and discussion}

\begin{figure}
\includegraphics[width=0.95\linewidth,clip=true]{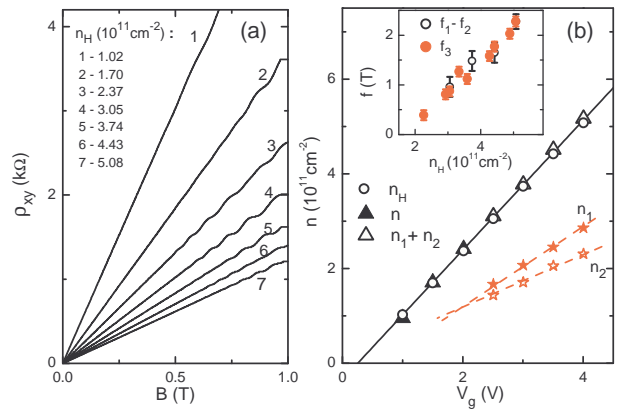}
\caption{(Color online) (a) -- The magnetic field dependences of $\rho_{xy}$ at various electron densities. (b) -- The gate voltage dependences of the Hall density,  $n_H=1/ eR_H (0.2\text{~T})$, and the density found from the Fourier spectra of the oscillations as $n_i=f_{i}\times e/h$, $i=1,2$ and $n=f_0\times 2e/h$. The inset shows the $n_H$ dependences of the difference frequency $f_1-f_2$ and the low-frequency component $f_3$ (see text). $T=4.2$ K.}
\label{F1}
\end{figure}

The experimental results and their analysis for all the  structures are similar, therefore, as an example, let us consideer in more detail the data obtained for the structure 100623 with $d=18$~nm. In Fig.~\ref{F1}(a), we present the dependences of transverse  magnetoresistance ($\rho_{xy}$) on a magnetic field ($B$) for different gate voltages. As seen, $\rho_ {xy}$ linearly increases with $B$ in low magnetic fields, then the  oscillations appear, which are transformed to the quantum Hall effect (QHE) steps in the higher magnetic field.

\begin{figure}
\includegraphics[width=\linewidth,clip=true]{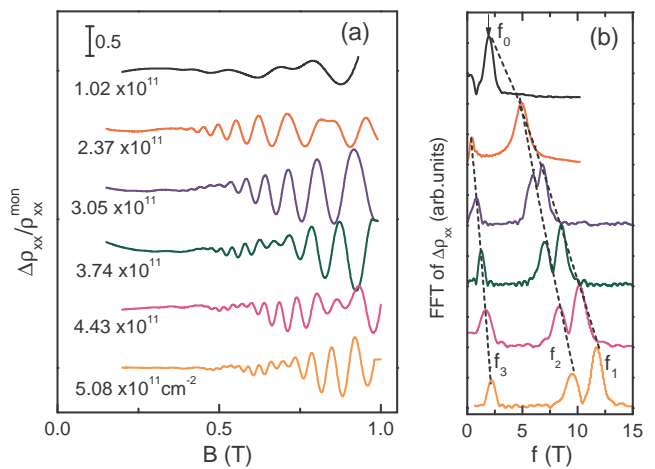}
\caption{(Color online) The magnetic field dependencies of the oscillating  part of magnetoresistance for some electron densities (a) and the Fourier spectra of these oscillations (b).
}
\label{F2}
\end{figure}

The $V_g$ dependence of the electron density obtained from the Hall effect as $n_H=-1/eR_H(0.2\text{~T})$  plotted in Fig.~\ref{F1}(b) shows that this dependence is linear, $n_H(V_g)= -0.35\times 10^{10}+dn/dV_g\times V_g$,~cm$^{-2}$ with $dn/dV_g=1.37\times 10^{11}$ cm$^{-2}$V$^{-1}$.  This $dn/dV_g$ value  is in a good agreement with $dn/dV_g$ found from the capacitance measurements $dn/dV_g=C/eS_g$, where $C$ is the capacitance between the 2D gas and the gate electrode, measured for the same structure, $S_g$ is the gated area. The oscillating part of magnetoresistance $\Delta\rho_{xx}(B)$ obtained by extracting the monotonic part $\rho_{xx}^{mon}(B)$, $\Delta\rho_{xx}(B)=\rho_{xx} (B)- \rho_{xx}^{mon}(B)$, for some gate voltages and  the Fourier spectra of these oscillations found in magnetic field range before onset of the QHE steps  are shown in Figs.~\ref{F2}(a) and Figs.~\ref{F2}(b), respectively. These data show that unsplit oscillations with one component in the Fourier spectrum  $f_0$ are observed for low electron density $n\lesssim 2.5\times 10^{11}$ cm$^{-2}$.

For the higher electron densities the beating of oscillations becomes clearly evident, the  $f_0$ component in the  Fourier spectrum  is split and two maxima, $f_1$ and$f_2$, arise in the Fourier spectra.  The electron densities found from $f_0$ supposing two-fold degeneracy of the Landau levels $n=f_0\times 2e/h$ are shown in Fig.~\ref{F1}(b) by the solid triangles.  The electron densities found from $f_1$ and$f_2$ supposing one-fold degeneracy of the Landau levels $n_i=f_{i}\times e/h$, $i=1,2$ are shown by the stars in Fig.~\ref{F1}(b) and the sum of these densities is shown by the open triangles. One can see that $n=f_0\times 2e/h$ and $n_1+n_2$ coincide with the Hall densities within experimental accuracy.
The above results show that the splitting of the SdH oscillations observed at $n>2.5\times 10^{11}$ cm$^{-2}$ is a consequence of the SO splitting of the spectrum, and $n_1$ and $n_2$ are the densities in the SO split subbands.

\begin{figure}
\includegraphics[width=0.95\linewidth,clip=true]{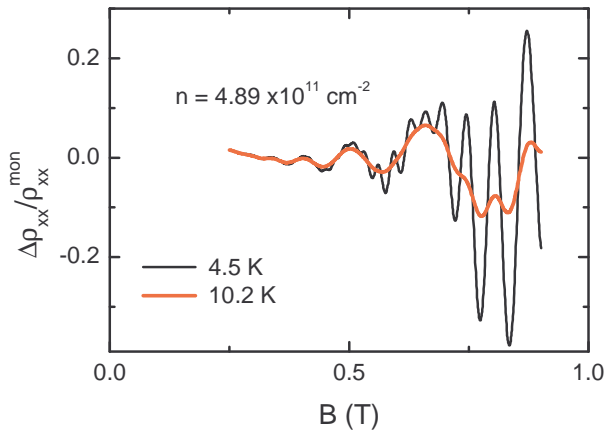}
\caption{(Color online) The oscillating  part of magnetoresistance for two temperatures. $n=4.89\times 10^{11}$ cm$^{-2}$V$^{-1}$. }
\label{F3}
\end{figure}

The origin of the  low-frequency oscillations that give the $f_3$ pick in the Fourier spectra becomes clear from the inset in Fig.~\ref{F1}(b), which demonstrates that $f_3$ is equal to $f_1-f_2$. This denotes that the low-frequency oscillations are a consequence of intersubband transitions, known as magneto-intersubband oscillations (MISO). This conclusion is  confirmed also by the temperature dependence of the amplitude of these oscillations, which, as the theory predicts \cite{Polyanovs88,Leadley88,Sander88,Rowe01}, should decrease slowly with the increasing temperature (see Fig.~\ref{F3}). A detailed analysis of MISO is beyond the scope of this paper.

As mentioned above the SO splitting can be characterized by the quantity   $\Delta n/n$. Experimentally, it can be found by two ways; as
\begin{equation}
\frac{\Delta n}{n}=\frac{n_1-n_2}{n_1+n_2} = \frac{f_1-f_2}{f_1+f_2}
\label{eq01}
\end{equation}
or as
\begin{equation}
\frac{\Delta n}{n} = \frac{e}{h}\frac{f_3}{n_1+n_2}=\frac{f_3}{f_1+f_2}.
\label{eq02}
\end{equation}

\begin{figure}
\includegraphics[width=0.95\linewidth,clip=true]{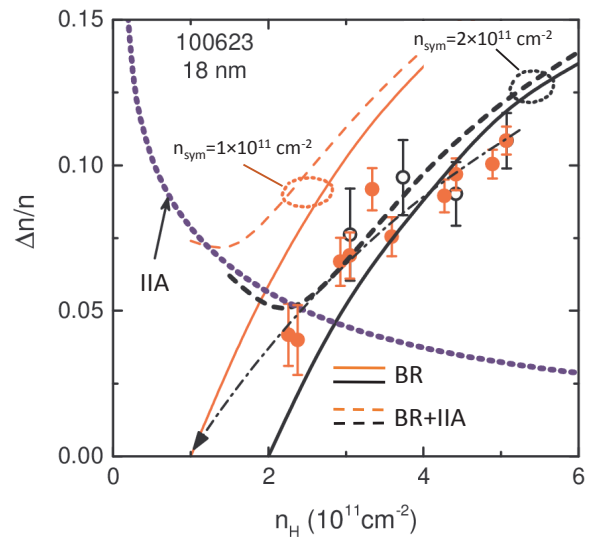}
\caption{(Color online) The experimental values of SO splitting found as $(f_1-f_2)/(f_1+f_2)$ at $T=4.2$~K (open circles) and as $f_3/(f_1+f_2)$ at $T=12$~K (solid circles)  for the various electron densities. The dash-dotted line  shows the extrapolation by eye of the data to $\Delta n/n=0$. Solid and dashed curves are the results of  theoretical calculations performed  without and with taking into account IIA, respectively, for different $n_{sym}$ (see the text). The doted curve is the theoretical dependence, which takes into account IIA only. The parameters from Ref.~\cite{Novik05} and $g_4=0.8$ have been used in the calculations.}
\label{F4}
\end{figure}

This approach has been used to obtain the SO splitting as a function of the electron density at relatively high electron density, $n>3\times 10^{11}$ cm$^{-2}$. The results are presented in Fig.~\ref{F4}. It can be seen that  the values of splitting  determined with the use of Eq.~(\ref{eq01}) and Eq.~(\ref{eq02})  agree well with each other. Therewith, the SO splitting  decreases with the decreasing electron density (with the decreasing gate voltage).

\begin{figure}
\includegraphics[width=0.95\linewidth,clip=true]{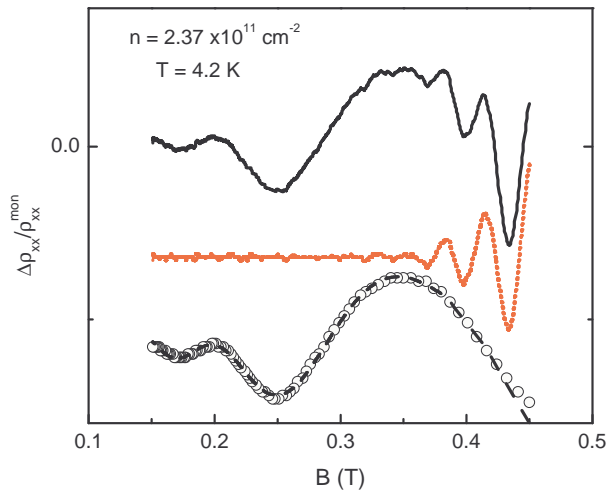}
\caption{(Color online)
\label{F5} From top to  bottom;  the oscillations of magnetoresistance measured at a low density, when the splitting of the $f_0$ Fourier component is not resolved (the upper curve) and decomposition result of the oscillations after the  Fourier filtering. The dashed line is  the result of the best fit of the low-frequency component by the Lifshitz-Kosevich-like formula. }
\end{figure}

At  $n<3\times 10^{11}$ cm$^{-2}$, the splitting of the $f_0$ peak ceases to be resolved (see Fig.~\ref{F2}). In this case, $\Delta n/n$ can only be determined by analyzing the  MISO plot  $\Delta\rho_{xx}(B)$. To extract MISO from the complicated experimental curve (see Fig.~\ref{F5}, the upper line) we  have decomposed the experimental oscillations into the low- and high frequency  components using the Fourier transformation and the low- and high-pass filtration. The frequency of the low-frequency oscillations  (i.e. MISO)  was determined from the fit by the Lifshitz-Kosevich-like formula, as demonstrated in Fig.~\ref{F5}.

\textbf{}

Now we are in position to compare the experimental results  with the theoretical calculations.  To do this  a self-consistent  calculation of spectrum was performed within the framework of the four-band \emph{kP} method which allows us to take into account both the interface inversion asymmetry and the BR effect (we used the parameters from Ref.~\cite{Novik05}) \footnote{The calculations were performed for three QW orientations corresponding to the structure under investigation. The difference in SO splitting $\Delta n/n$ does not exceed $10$~percent}. For such a calculation, it is necessary to specify the electron density  ($n_{sym}$) at which the electric potential in the quantum well is symmetric. As the first step, let us determine  this value by eye, extrapolating the experimental data to  zero $\Delta n/n$ value as shown by the dash-dotted line in Fig.~\ref{F4}. Doing so we get $n_{sym}=1\times 10^{11}$ cm$^{-2}$. The results of the calculation without and with taking into account IIA are shown in Fig.~\ref{F4} by the upper solid and dashed curves, respectively  \footnote{We used $g_4 = 0.8$ eV$\times{\AA}$ \cite{Minkov17}. This value gives a good agreement with the experimental splitting values of the top of the valence band in the structures with $d\simeq d_c$, in which the top of the valence band is located at $k=0$ \cite{Minkov16}}. It can be seen that these dependences lie noticeably higher than the experimental values. The same figure shows the calculation results for the other $n_{sym}$ value, $n_{sym}=2\times 10^{11}$ cm$^{-2}$ (the lower solid and dashed curves in Fig.~\ref{F4}). In this case, as seen, both curves  describe the experimental data equally well. For clarity, we present by the doted curve the dependence calculated with taking into account the IIA only. It is apparent that IIA gives a large contribution to $\Delta n/n$ at the electron density close to $n-n_{sym} <1\times 10^{11}$ cm$^{-2}$ and a small contribution at the higher density, where practically all the experimental points lie. We would like to note here that the additive sum of curves, calculated with taking into account only the interface inversion asymmetry or the BR effect separately does not coincide with the curve calculated with taking into account both mechanisms.

Thus, analysis of the results obtained on the structure with the  QW width of $18$~nm shows that the Bychkov-Rashba effect is the main cause of the SO splitting, while the IIA contribution to SO splitting  is small and cannot be determined quantitatively.

\begin{figure*}
\includegraphics[width=0.95\linewidth,clip=true]{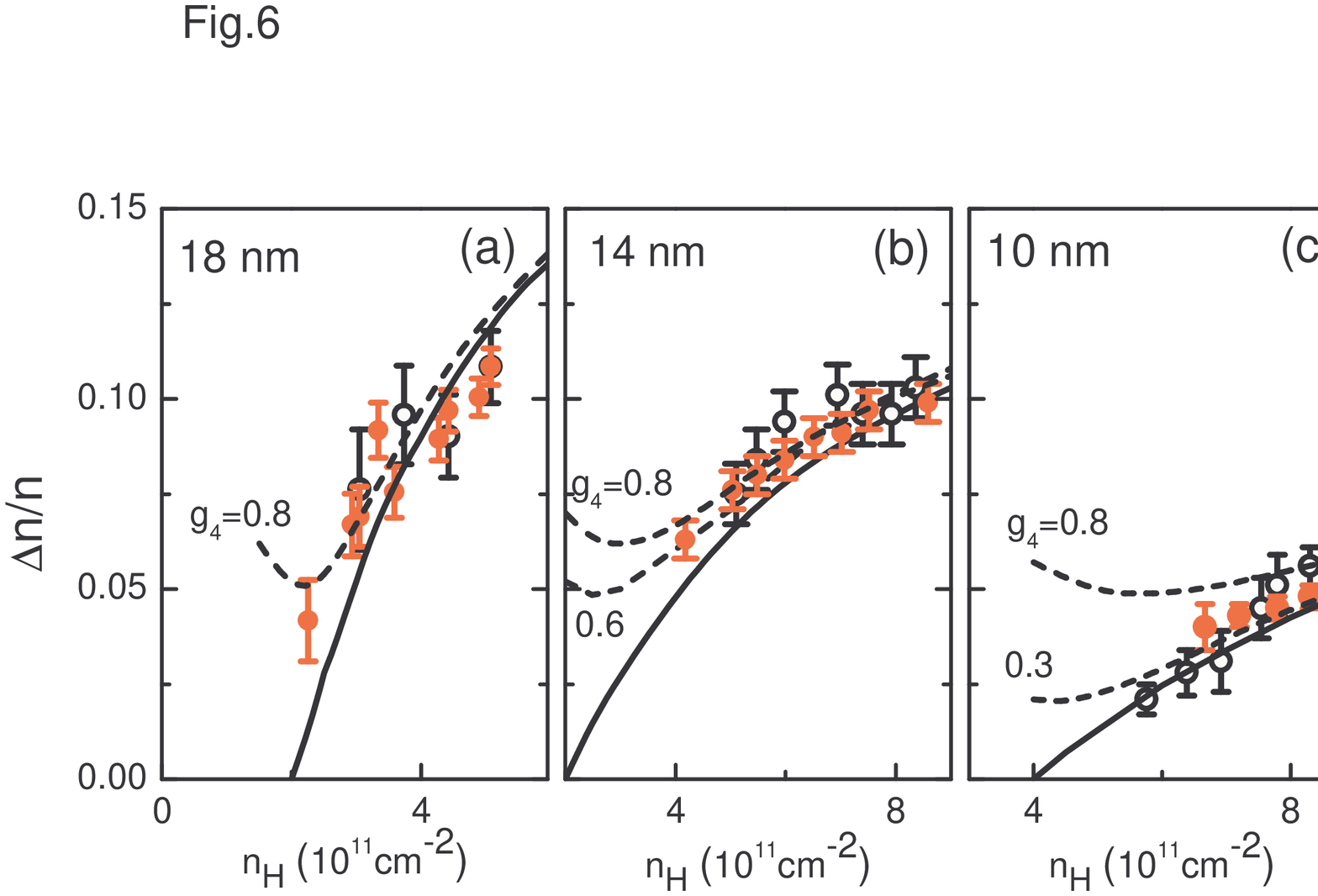}
\caption{(Color online) The electron density dependences of SO splitting for all the structures from the Table~\ref{tab1}. The symbols are the experiment. The solid and dashed curves are the calculation result without and with taking into account IIA, respectively. }
\label{F6}
\end{figure*}

The above measurements and analysis  were carried out for all the structures listed in the Table~\ref{tab1}. The results are summarized in Fig.~\ref{F6}. It is seen that as the QW width decreases, both the value and the rate of change of $\Delta n/n$ decrease significantly, so it is more difficult to determine $\Delta n/n$.  Nevertheless, this can be done in the structure with $d=8.3$~nm. Note that for $(n-n_{sym})<4.6\times 10^{11}$ cm$^{-2}$, the $\Delta n/n$ value can only be determined from еру low-frequency oscillations, i.e. from MISO. The separation of MISO and determination of $f_3$ was analogous to that described above for the structure with $d=18$~nm (see Fig.~\ref{F5}).

Figure~\ref{F6} shows as well the results of calculating $\Delta n/n$, similar to those described above, for the structure 100623 with $d=18$~nm. It can be seen that taking into account only the BR mechanism  allows us to describe the experimental dependences for all the structures well (see the solid curves in Fig.~\ref{F6}). Satisfactory agreement of the theory with the data can also be obtained with an additional taking into account of IIA (see the dashed curves in Fig.~\ref{F6}). However, we must monotonically reduce the upper value of the parameter $g_4 $ with decreasing  QW width.

These results show that the contribution of IIA to SO splitting is small in comparison with the BR effect, which makes it impossible to quantify its contribution. Only in the structure 150629 with $d=8.3$~nm, accounting for IIA gives a somewhat better agreement with the experiment [see Fig.~\ref{F6}(d)]. Therewith, the $g_4$ value corresponding to the best agreement,  $g_4 = 0.2$, turns out to be four times lower than it was required in the analysis of the splitting of the valence band in Ref.~\cite{Minkov17}.
This result seems strange. The matter is that the SO splitting due to IIA is a result of different mixing of the states at the opposite QW interfaces. This mixing  is described   by the same parameter $g_4$ both for the valence and for conduction band within the framework of the model used. From our point of view, this shows that the usually used theoretical model does not take into account something important.

Let us discuss what approximations were made when analyzing the data. First, the use of Eq.~(\ref{eq01}) for the  determination of $\Delta n/n$ means that we neglect the contribution of the Zeeman splitting in comparison with the spin-orbit splitting. Simultaneous accounting for the Zeeman and SO splitting leads to more complicated expression for the magnetic field dependence of the Landau level energies \cite{Winkler} from which it follows that the oscillation frequencies become dependent on the magnetic field because the Zeeman splitting increases linearly with $B$ increase while the SO splitting is independent of  magnetic field. Our estimations show that neglect of the Zeeman splitting when treating the data at $B<(0.5-0.6)$~T  results in error in determination of $\Delta n/n$ less than $(10-15)$ percent for the our electron density range.

Second, we do not take into account the Dresselhaus contribution to SO splitting in our theoretical considerations. Parameters responsible for this contribution are unknown for HgTe and Hg$_{1-x}$Cd$_x$Te of actual composition. If one uses the parameters for CdTe \cite{Winkler},   this contribution is estimated as  ten-to-twenty times less than the observed values of $\Delta n/n$.

Finally,   we show in Ref.~\cite{Minkov18} that the experimental electron effective mass is not described within the framework of the  \emph{kP} model in its usual form. We assumed there that this is result of renormalization of the spectrum of the conduction band due to many-body effects. When analyzing SO interaction in the present paper, we also used the  \emph{kP} calculation in its conventional form. We do not know the papers in which the SO interaction is considered taking into account such renormalization. Recall  that we were able to estimate the upper value of the IIA contribution from the data for structure with $d=8.3$~nm. The experimental effective mass obtained for this structure is only $15$ percent lower than the calculated value. Therefore we assume that the use of the \emph{kP} calculations for our case is reasonable.

\section{Conclusion}
We have presented the results of experimental studies of spin-orbit splitting of the conduction band in HgTe quantum wells. In order to elucidate  the role of different mechanisms of SO interaction, we carried out detailed measurements of the SdH oscillations in structures with quantum well widths from $8$ to $18$~nm over a wide range of electron density. We observed appearance of three components in the Fourier spectra characterized by the frequencies  $f_1$, $f_2$, and $f_3$  when electron density is increased by the gate voltage. The analysis of these results shows that the $f_1$ and $f_2$ components  give the electron densities $n_1$ and $n_2$ in subbands split by SO interaction, while the $f_3$ component  results from the  magneto-intersubband oscillations and it is determined by the difference between the electron densities in SO split subbands $\Delta n$. Treating the Fourier spectra with taking this into account, we have determined the value of the SO splitting $\Delta n/n$ over the wide electron density range for all the structures under study. The data obtained were compared with the results of self-consistent calculations carried out within the framework of the four-band $kP$-model. It is  shown that the main contribution to the SO splitting comes from the Bychkov-Rashba effect while the contribution of the interface inverse asymmetry is less than $(10-15)$ percent. Thus we have obtained surprising  results. In HgTe-QWs, the interface inverse asymmetry  is not so important in the SO splitting of the conduction band unlike the valence band, where it plays a crucial role as shown in Refs.~\cite{Minkov16,Minkov17}.

The further experimental and theoretical
investigations are needed to solve this puzzle.

\acknowledgements

We are grateful to I.~V.~Gornyi and L.~E.~Golub for useful discussions.
The work has been supported in part by the Russian Foundation for Basic
Research (Grants No. 16-02-00516 and No. 18-02-00050), by  Act 211 Government of the Russian Federation, agreement No.~02.A03.21.0006,  by  the Ministry of Education and Science of the Russian Federation under Project No.~3.9534.2017/8.9, and by the FASO of Russia (theme ``Electron'').


%

\end{document}